\documentclass{aa}
\usepackage{graphicx}
\usepackage{txfonts}
\begin{document}
   \title{A pair of CO + He white dwarfs as the progenitor of 2005E-like supernovae?}


   \author{Xiangcun Meng
          \inst{1,2}
          and
          Zhanwen Han \inst{1,2}}

   \offprints{X. Meng \& Z. Han}

   \institute{$^{\rm 1}$Yunnan
Observatories, Chinese Academy of Sciences, Kunming, 650216,
China\\
              \email{xiangcunmeng@ynao.ac.cn, zhanwenhan@ynao.ac.cn}\\
$^{\rm 2}$Key Laboratory for the Structure and Evolution of
Celestial
Objects, Chinese Academy of Sciences, Kunming, 650216, China\\
             }

   \date{Received; accepted}


  \abstract
   {Ca-rich transients (CRTs, 2005E-like supernovae) exhibit unusually strong Ca features during their nebular phase,
   and their distribution in their host galaxies indicates that they belong to a metal-poor old
   population. A pair of low-mass CO + He white dwarfs (WD) has been suggested
   to be the progenitor of CRTs. A helium shell is accumulated onto the CO WD by accretion, and
   then a
   helium-shell detonation is ignited
   when the helium shell reaches a critical mass, which could lead to the second detonation in
   the center of the CO WD.
   }
   {Taking the birth rate of CRTs into consideration, we examine whether and if yes,
    which type of low-mass CO + He WD pairs
   fulfill the constraints of being of an old population and of the
   birth rate derived from observations.
}
   {We carried out a series of binary population syntheses and
   present four different channels in which CO + He WD
   pairs can be formed. We selected the systems that fulfill the constraints of being of an old population
   and of the
   birth rate from all the CO + He WD
   pairs by constraining the component mass of the WD pairs.
}
   {For the four channels, the stable Roche lobe overflow (RLOF) could significantly influence the formation of the WD
   pairs. Based on their position on the $M_{\rm CO}$-$M_{\rm He}$
   plane, the mass-transfer between the components for most of the CO + He WD pairs is neither always unstable nor always stable.
   We found that it is necessary that the CO WDs are less massive than 0.6 $M_{\odot}$ and
   the He WDs are less massive than 0.25 $M_{\odot}$ if CO + He WD pairs are to fulfill the
   constraints of being of an old population and of the birth rate of
   CRTs. However, the He WD mass is lower than the ejecta
   mass of the CRTs derived from observations, while the total mass of the low-mass WD pairs is higher than this.}
   {Our results imply that the CO WDs
   participate in CRT explosions and at the same time, a bound remnant could be left after the CRT explosion if the low-mass WD pairs are the progenitors of
   CRTs. Therefore, it needs to be examined whether or not the helium detonation on a low-mass CO
   WD may lead to the second detonation in the center of the CO WDs, and whether or not the bound remnant is left after the thermonuclear
explosion.}

   \keywords{Stars: white dwarfs - stars: supernova:
   general-individual: SN 2005E
               }
   \authorrunning{Meng \& Han}
   \titlerunning{A pair of CO + He WDs as the progenitor of 2005E-like supernovae?}
   \maketitle{}
%

\section{Introduction}\label{sect:1}
Supernovae (SNe) are one of the possible evolutionary end stages
of stars and are divided into several observational subclasses
according to the line features of different chemical elements in
their spectrum. Type Ia supernovae (SNe Ia) generally result from
the thermonuclear explosion of a white dwarf (WD) in a binary
system (Hillebrandt \& Niemeyer \cite{HN00}; Leibundgut
\cite{LEI00}), while Type Ib/c supernovae (SNe Ib/c) and Type II
supernovae (SNe II) result from the gravitational core-collapse of
a star of $\geq 8$ $M_{\odot}$ (Smartt \cite{SMARTT09}). However,
the discovery of SN 2005E appears to contradict this
understanding, because this supernova was classified as a SN Ib
while it seems impossible for it to have arisen from a massive
star. This means that it probably belongs to a new type of stellar
explosion (Perets et al. \cite{PERETS10}). The SNe belonging to
this new subclass share many common properties: 1) a low peak
luminosity with a magnitude between that of regular novae and
supernovae ($M_{\rm R}=-16\pm0.5$ mag); 2) a relatively fast rise
and decay time ($\sim12-15$ days); 3) a photospheric velocity
similar to normal SNe Ia; 4) a quick spectral transition into the
nebular phase ($\sim1-3$ months); 5) peculiar nebular spectra
dominated by calcium lines; 6) They are located in border regions
of host galaxies and then belong to a metal-poor old population
(even older than 10 Gyr); and they have 7) fewer ejecta in terms
of mass (0.4-0.7 $M_{\odot}$) (Kasliwa et al. \cite{KASLIWA12};
Yuan et al. \cite{YUANF13}; Lyman et al. \cite{LYMAN13}). As a
result of the dominant calcium lines in the nebular phase, these
SNe are called Ca-rich transients (CRTs), and the sample size of
this subclass is still increasing (Valenti et al.
\cite{VALENTI14}). Perets et al. (\cite{PERETS10}) estimated that
the birth rate of CRTs is $7\%\pm5\%$ relative to the total SNe Ia
rate (see also Kasliwa et al. \cite{KASLIWA12}). At present, the
identity of CRT progenitors is still unclear (Kawabata et al.
\cite{KAWABATA10}; Perets et al. \cite{PERETS11}). Considering
that the CRTs originate from a metal-poor old population and that
some CRTs exhibit helium lines in their spectra, a binary with a
carbon-oxygen white dwarf (CO WD) primary and a helium-rich
secondary was proposed (Perets et al. \cite{PERETS10}), and the
helium-rich secondary is more likely to be a helium (He) WD since
a helium star usually belongs to a young population (Kasliwa et
al. \cite{KASLIWA12}; Yuan et al. \cite{YUANF13}; Lyman et al.
\cite{LYMAN13}). Observationally, there are indeed such WD pairs,
and the mass-transfer between the components of some WD pairs may
be dynamically stable, which could mean that the accreted helium
may be gradually accumulated onto the CO WD, eventually leading to
a helium-shell detonation (Kilic et al. \cite{KILIC12,KILIC14}).
In theory, a helium-shell detonation on a low-mass CO WD, where
the helium-rich material is accreted from a low-mass He WD, may
reproduce the properties of the prototype of the CRTs (SN 2005E),
while for a massive CO WD, the detonation  mainly leaves $^{\rm
56}$Ni and unburnt helium, and the predicted spectrum is unlikely
to fit the unique features of CRTs (Shen et al. \cite{SHENK10};
Waldman et al. \cite{WALDMAN11}). Although observations and
numerical simulations support CO + He WD pairs as the progenitor
of CRTs, it is still unclear which type of CO + He WD pairs may
contribute to the CRTs and whether the CO + He WD pairs can
account for the delay time and birth rate of CRTs. It is
especially unclear whether the CO + He WD pairs can account for
other properties of CRTs such as the ejecta mass. Judging from
observational constraints, the WD pairs producing CRTs should
belong to metal-poor old populations, and their merging rate
should be consistent with the birth rate of CRTs. In this paper,
we try to discuss these problems based on a detailed binary
population synthesis (BPS) study.

In Sect. \ref{sect:2}, we describe our BPS method, and we present
the results in Sect. \ref{sect:3}. We provide our discussions in
Sect. \ref{sect:4} and our conclusions in Sect. \ref{sect:5}.

   \begin{figure*}
   \centering
   \includegraphics[width=190mm,height=130mm,angle=0.0]{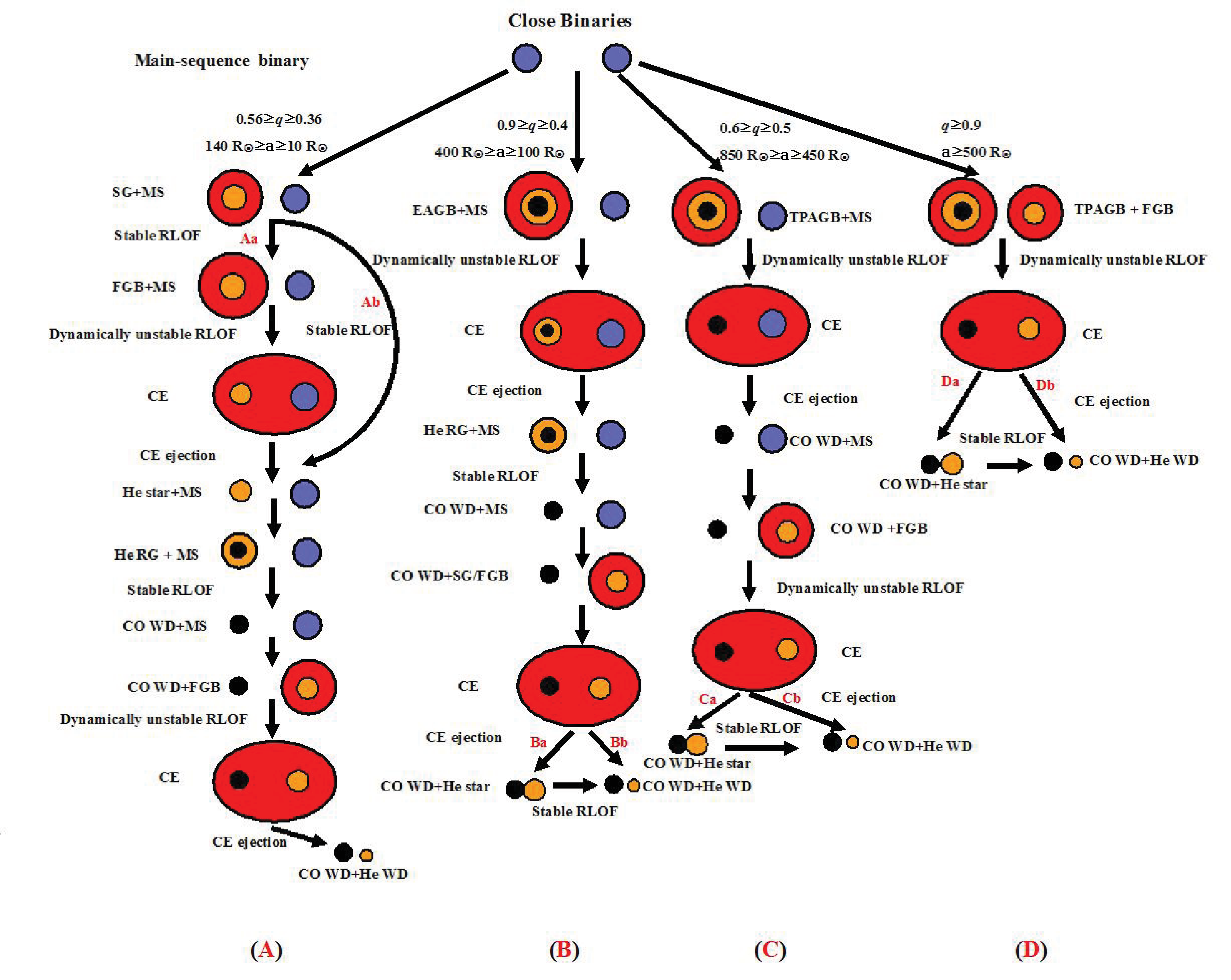}
   \caption{Schematic for the evolution channels that can form the CO WD + He WD pairs, where $q=M_{\rm 2}/M_{\rm 1}$
   is the primordial mass ratio and $a$ is the primordial orbital separation (see text for details).} \label{Fig1}%
    \end{figure*}


\section{Binary population synthesis method}\label{sect:2}

A very complex process is required to form a CO + He WD pair, and
different primordial systems will experience different evolution
channels, in which they will undergo one or two common-envelope
(CE) phases. We first introduce the different evolution channels
in which the CO + He WD pairs can form, and the methods with which
we treat the CE evolution in this chapter.

\subsection{Evolution channel}\label{sect:2.1}
From a primordial main-sequence (MS) binary, a CO + He WD pair may
form on an evolutionary timescale $t_{\rm E}$ after one or two CE
phases (Han \cite{HAN98}; Nelemans et al. \cite{NELEMANS05};
Ruiter et al. \cite{RUITER09}). $t_{\rm E}$ is mainly determined
by the evolutionary timescale of the secondary. Following the CO +
He WD pair, orbital angular momentum loss by gravitational-wave
(GW) radiation dominates the evolution of the system, and finally,
the system merges within a timescale $t_{\rm GW}$ (Landau \&
Lifshitz \cite{LANDAU62}),
\begin{equation}
t_{\rm GW}{\rm (yr)}= 8\times10^{\rm 7}\times\frac{(M_{\rm
1}+M_{\rm 2})^{\rm 1/3}}{M_{\rm 1}M_{\rm 2}}P^{8/3},\label{eq:gw}
  \end{equation}
where $P$ is the orbital period of the pair in hours, and $M_{\rm
1}$ and $M_{\rm 2}$ are the masses of the CO WD and the He WD in
solar mass, respectively. Thus, the delay time elapsed from the
birth of a primordial binary system to the occurrence of the
merger, $t_{\rm tot}$, is equal to the sum of $t_{\rm E}$ and
$t_{\rm GW}$. We assume that if $t_{\rm tot}$ is shorter than 15
Gyr, the merger is a potential candidate to be the progenitor of
the CRTs. The delay time $t_{\rm tot}$ can also be used to
determine whether the progenitor of the CRTs belongs to a young or
old population. We identify the progenitors from these potential
candidates, based on the constraints of the old population and
birth rate of the CRTs derived from observations.

Based on the number of the CEs that the WD pairs experienced and
the time that the primordial primary fills its Roche lobe, there
are four channels in which the CO + He WD pairs can form from
primordial binaries, as shown in Fig. \ref{Fig1}. Below, we
briefly outline the four channels: Cases A, B, C, and D.

Case A (2RLOF+1/2CE): The primordial zero-age main-sequence (ZAMS)
mass of primaries is relatively massive: 2.5 -- 4.5 $M_{\odot}$,
while that of the secondary is between 0.9 -- 2.5 $M_{\odot}$, and
the primordial orbital separation is short, 10 -- 140 $R_{\odot}$.
Case A is divided into two subchannels, Cases Aa and Ab. The main
difference between the two subcategories lies in different
primordial orbital separations, which determine the time of the
first Roche-lobe overflow (RLOF) and whether a CE forms following
the first RLOF. For Case Aa, the first RLOF occurs when the
primary is at the end of the Hertzprung gap (HG). After a short
period of stable RLOF, the primary becomes a first giant branch
(FGB) star, and then the RLOF becomes dynamically unstable,
leading to the formation of a CE. After the CE ejection, the
primary becomes a helium star and continues to evolve. The helium
star fills its Roche lobe (RL) again after exhausting the central
helium, and the second stable RLOF begins at a relatively low mass
ratio, which leads to the formation of a CO WD + MS system.
Following this, the secondary evolves to the FGB and the RLOF is
also unstable, resulting in the second CE. After ejecting the CE,
a CO + He WD pair forms. For Case Ab, the first RLOF may always be
stable if the primary fills its RL when it is crossing the HG.
Thus, instability may be avoidable for the first CE, and the
system becomes a He star + MS system directly after the first RLOF
(Han et al. \cite{HAN02}). The following evolution is similar to
Case Aa. After the merging of the WD pairs, Case A mainly
contributes to the young and medium-age population

Case B (2CE+1/2RLOF):  This channel is also divided into two
subchannels, Cases Ba and Bb. The primordial mass ratio ($m_{\rm
2}/m_{\rm 1}$) is between 0.4 and 0.9, and the primordial
separation is mainly from 100 $R_{\odot}$ to 400 $R_{\odot}$ for
both subchannels. For Case Ba, the primordial ZAMS primary is
relatively massive, 3 -- 4.5 $M_{\odot}$, while it is smaller, 1.6
-- 3.3 $M_{\odot}$, for Case Bb. The primary fills its RL when it
is an early asymptotic giant branch (EAGB) star. The following
RLOF is dynamically unstable for a deep convective envelope, which
results in a CE. After the CE ejection, the primary becomes a
helium red giant (RG) star. Soon, the helium RG star fills its RL
again, and the following RLOF is dynamically stable for a low mass
ratio. Then, a CO WD + MS forms after the RLOF. When the secondary
becomes an FGB star, it fills its RL and the RLOF is dynamically
unstable, leading to a CE. The surviving systems resulting from
cases Ba and Bb are different. For Case Bb, a CO + He WD pair
forms directly after the CE ejection, while for Case Ba it is a CO
WD + He star system because of the relative high mass of the
secondary. For the CO WD +He star system, the orbital separation
is so small for the two CE evolution that the He star fills its RL
during the helium-burning phase of its main sequence. When the
mass of the helium star is lower than the lowest value required to
maintain helium burning, the helium burning is quenched in the
center of the helium star, and the helium star becomes a He WD
(Han et al. \cite{HAN02}). A CO + He WD pair forms as a result.
After the merging of the system, Case Ba mainly contributes to
merger formation in young populations, while Case Bb contributes
to that of the medium-age and old populations.

Case C (2CE+0/1RLOF): this channel is divided into two
sub-channels, Cases Ca and Cb. The primordial mass ratio ($m_{\rm
2}/m_{\rm 1}$) is between 0.5 and 0.6, and the primordial
separation ranges from 450 $R_{\odot}$ to 850 $R_{\odot}$ for both
subchannels. For Case Cb, the mass of the primordial ZAMS primary
is from 2$M_{\odot}$ to 4 $M_{\odot}$, while it is usually more
massive than 4 $M_{\odot}$ for Case Ca. The primary fills its RL
when it is a thermal pulsing asymptotic giant branch (TPAGB) star.
The RLOF is then dynamically unstable for a deep convective
envelope, leading to a CE. After CE ejection, a CO WD + MS system
forms. The secondary fills its RL when it is climbing the FGB, and
then the following RLOF also becomes dynamically unstable,
resulting in a second CE. After the ejection of the second CE, the
surviving system directly becomes a CO + He WD pair for Case Cb
and a CO WD + He star for Case Ca. The following evolution of Case
Ca is similar to Case Ba. After the merging of the WD pairs, Case
Ca mainly contributes to merger formation in young populations,
while Case Cb contributes to that of the medium-age and old
populations.

Case D (1CE+0/1RLOF): This channel is also divided into two
sub-channels, Cases Da and Db. The primordial mass ratio ($m_{\rm
2}/m_{\rm 1}$) is higher than 0.9, even close to 1 in some cases,
and the primordial separation is from 500 $R_{\odot}$ to 1000
$R_{\odot}$ for Case Db, while it can be larger than 1000
$R_{\odot}$ for Case Da. The mass of the primordial primary is
from 1 $M_{\odot}$ to 2.5 $M_{\odot}$. As a result of the high
mass ratio and the large primordial orbital separation, the
secondary is an FGB star when the primary fills its RL at the
TPAGB stage. The following RLOF results in a CE, where the CE also
includes the envelope of the FGB star, and the binary embedded in
the CE consists of the dense cores of the TPAGB and the FGB stars.
After the CE ejection, the surviving system is a CO + He WD pair
for Case Db, while a CO WD + He star results from Case Da. The
following evolution of Case Da is similar to that of Cases Ba and
Ca. After the merging of the WD pairs, Case Da mainly contributes
to merger formation in medium-age populations, while Case Db
contributes to that of old populations.

\subsection{Common envelope}\label{sect:2.2}
As mentioned above, the CE phase is very important for the
formation of the CO + He WD pairs. The mass ratio ($q=M_{\rm
donor}/M_{\rm accretor}$) is crucial during  binary evolution. If
the mass ratio is higher than a critical value, $q_{\rm c}$, mass
transfer between the two components is dynamically unstable, and a
CE will form (Paczy$\acute{\rm n}$ski \cite{PAC76}). The critical
ratio $q_{\rm c}$ varies with the evolutionary state of the donor
star at the onset of the RLOF (Hjellming \& Webbink\cite{HW87};
Webbink \cite{WEBBINK88}; Han et al. \cite{HAN02}; Podsiadlowski
et al. \cite{POD02}; Chen \& Han \cite{CHE08}). We adopted $q_{\rm
c}$ = 4.0 when the donor star is a MS star or is in the HG,
following previous detailed binary evolution studies (Han et al.
\cite{HAN00}; Chen \& Han \cite{CHE02, CHE03}). If the primordial
primary is on the FGB or the AGB, we adopted
\begin{equation} q_{\rm c}=[1.67-x+2(\frac{M_{\rm c}}{M})^{\rm 5}]/2.13,  \label{eq:qc}
  \end{equation}
where $M_{\rm c}$ is the core mass of the donor star and $x={\rm
d}\ln R_{\rm 1}/{\rm d}\ln M$ is the mass--radius exponent of the
donor star and varies with composition. If the mass donors
(primaries) are naked helium giants, $q_{\rm c}$ = 0.748, as
calculated from Eq. (\ref{eq:qc}) (see Hurley et al. \cite{HUR02}
for details).

The binary embedded in the CE consists of the dense core of the
donor star and the secondary. The  orbit of the embedded binary
gradually shrinks owing to frictional drag introduced by the
envelope, and a large part of the orbital energy released during
the spiral-in process is injected into the CE (Livio \& Soker
\cite{LS88}). Here, we assumed that the CE is ejected if
\begin{equation}
\alpha_{\rm CE}\Delta E_{\rm orb}\geq|E_{\rm bind}|,
\label{eq:alpha}
  \end{equation}
where $\Delta E_{\rm orb}$ is the orbital energy released during
the spiral-in process, and $E_{\rm bind}$ is the binding energy of
the CE. It is very sensitive to the value of the CE ejection
efficiency, $\alpha_{\rm CE}$, which represents the fraction of
the released orbital energy used to eject the CE. Since the
thermal energy in the envelope is not incorporated into the
binding energy here, it is possible for $\alpha_{\rm CE}$ to be
greater than 1 (see Han et al. \cite{HAN95} for details about the
thermal energy). At present, the value of $\alpha_{\rm CE}$ is
very uncertain (Zorotovic et al. \cite{ZOROTOVIC10}; Camacho et
al. \cite{CAMACHO14}; Zuo \& Li \cite{ZUO14}). We here set
$\alpha_{\rm CE}=3.0$ as our standard value and $\alpha_{\rm
CE}=1.0$ to test the influence of $\alpha_{\rm CE}$ on the final
results.

After the CE evolution, the initial orbital separation at the
onset of the CE phase, $a_{\rm i}$, becomes $a_{\rm f}$, which is
determined by
\begin{equation}
\frac{G(M_{\rm c}+M_{\rm e})M_{\rm e}}{\lambda R_{\rm
1}}=\alpha_{\rm CE}(\frac{GM_{\rm c}m}{2a_{\rm
f}}-\frac{GMm}{2a_{\rm i}}),
  \end{equation}
where $\lambda$ is a structural parameter depending on the
evolutionary stage of the donor. $M$, $M_{\rm c}$, and $M_{\rm e}$
are the masses of the donor, the donor core, and the envelope,
respectively. $R_{\rm 1}$ is the radius of the donor, and $m$ is
the mass of the secondary. We assumed a constant structural
parameter and set $\lambda=0.25$, $0.5$ and $1.0$ to test its
effect on the final results (de Kool, van den Heuvel \& Pylyser
\cite{DEKOOL87}), since a variable $\lambda$ may not significantly
affect the final results compared with a constant value of
$\lambda$, such as $\lambda=1.0$ (Claeys et al. \cite{CLAEYS14}),
although an exact calculation should take into account that
$\lambda$ depends on the stellar structure\footnote{Different
authors showed different formulae for $\lambda$, which means that
there are many uncertainties in calculating the value of
$\lambda$, such as the definition of the core of an AGB star and
the stellar chemical composition (Xu \& Li \cite{XL10}; Claeys et
al. \cite{CLAEYS14}). In addition, our code is an old version and
the subroutine for calculating the variable $\lambda$ is not
included.}. Thus, the final orbital separation of a surviving
binary system after the CE phase, $a_{\rm f}$, is determined by

\begin{equation}
\frac{a_{\rm f}}{a_{\rm i}}=\frac{M_{\rm c}}{M}(1+\frac{2M_{\rm
e}a_{\rm i}}{\alpha_{\rm CE}\lambda mR_{\rm 1}})^{\rm -1}.
\label{eq:af}
  \end{equation}
Here, we may combine $\alpha_{\rm CE}$ and $\lambda$ into one free
parameter $\alpha_{\rm CE}\lambda$, and then its value is $0.25$,
$0.5$, $1.0$,  or $1.5$ based on the selected value of
$\alpha_{\rm CE}$ and $\lambda$ (e.g., L\"{u} et al.
\cite{LUGL06}). From Eq. (\ref{eq:af}), we can see that the
results from simulations with $\alpha_{\rm CE}=1.0$ and
$\lambda=0.25$ or $0.5$ are exactly the same as those with
$\alpha_{\rm CE}=0.5$ and $\lambda=0.5$ or $1.0$.

Nelemans et al. (\cite{NELEMANS00}) and Nelemans \& Tout
(\cite{NELEMANS05}) argued that it may be difficult for this
prescription to produce a close pair of white dwarfs and suggested
an alternative algorithm that conserves angular momentum. However,
the feasibility of this suggestion is itself open to debate (see
the reviews by Webbink \cite{WEBBINK08}; Ivanova \cite{IVANOVA11};
Woods et al. \cite{WOODS11}). Especially when one tries to
constrain the CE mechanism by WD + WD pairs observed, it is
commonly assumed that the WD + WD pairs only experienced two CE
phases (Case Cb here) and the influence of RLOF is neglected
(Woods et al. \cite{WOODS11}). However, as described above, there
are four channels by which CO + He WD pairs can be formed, and
subchannel Cb is not dominant. As a result of these
considerations, the impact of RLOFs on the formation of WD + WD
pairs warrants further attention, and we did not implement the
suggestion of Nelemans et al. (\cite{NELEMANS00}) and Nelemans \&
Tout (\cite{NELEMANS05}) here.

\subsection{Basic parameters for Monte Carlo simulations}\label{sect:2.3}
To investigate the birth rate of CRTs from CO + He WD pairs, we
carried out a series of detailed Monte Carlo simulations with the
Hurley rapid binary evolution code (Hurley et al.
\cite{HUR00,HUR02}). We followed the evolution of $10^{\rm 7}$
sample binaries until a CO + He WD pair formed. The $10^{\rm 7}$
sample binaries were generated by a Monte Carlo algorithm based on
the following assumptions: (1) all the stars form in a single
starburst (where $10^{\rm 11} M_{\odot}$ of stars are formed at
the same instant); (2) the mass distribution of the newly formed
stars follows the initial mass function (IMF) of Miller \& Scalo
(\cite{MS79}); (3) the binaries either follow a constant
mass-ratio distribution, that is, $n(q)=1$, a rising distribution,
that is, $n(q)=2q$, or both binary components are chosen randomly
and independently from the same IMF (uncorrelated); (4) the binary
separation distribution is constant in $\log a$ for wide binaries,
where $a$ is the orbital separation; (5) binary orbits are either
circular or follow a rising distribution, that is, $n(e)=2e$; (6)
All stars have a metallicity of $Z=0.001$ (see Table \ref{Tab:1}
for details). We also simulated a case with solar metallicity to
test its effects, even though CRTs typically favor a
low-metallicity environment (Yuan et al. \cite{YUANF13}).

Based on these simulations, we calculated the evolution of the
birth rate of the merger under various conditions. For a merger to
produce a CRT, it should satisfy two terms: 1) the merger must
have a long delay time; 2) the birth rate of the merger should
match the birth rate of observed CRTs within error margins at a
delay time longer than 10 Gyr. We arbitrarily chose 1 Gyr and 3
Gyr to be loose and strict age boundaries, respectively, for the
purpose of judging whether a population is old. Most mergers
probably have a delay time longer than this boundary, since,
statistically, the ratio of the number of young CRTs to that of
all CRTs must be very low. In addition, according to the
estimation of Perets et al. (\cite{PERETS10}), the birth rate of
CRTs at a delay time of longer than 10 Gyr is normalized to the
birth rate of the SNe Ia in S0/E galaxies (Mannucci et al.
\cite{MANNUCCI05})\footnote{The birth rate of SNe Ia in S0/E
galaxies in Li et al. (\cite{LIWD11}) is slightly higher than that
in Mannucci et al. (\cite{MANNUCCI05}), but the difference does
not affect our final conclusion.}.

\begin{table}[]
\caption[]{Parameters used to produce the binary sample.}
\label{Tab:1}
\begin{center}
\begin{tabular}{ccccc}
\hline\noalign{\smallskip}
 set  & $\alpha_{\rm CE}\lambda$ & $n(q)$        & ecc      & Z\\
\hline\noalign{\smallskip}
 1    & 0.25              & constant      & circular & 0.001\\
 2    & 0.5               & constant      & circular & 0.001\\
 3    & 1.0               & constant      & circular & 0.001\\
 4    & 1.5               & uncorrelated  & circular & 0.001\\
 5    & 1.5               & constant      & circular & 0.001\\
 6    & 1.5               & $2q$          & circular & 0.001\\
 7    & 1.5               & constant      & $2e$     & 0.001\\
 8    & 1.5               & constant      & circular & 0.02\\
\noalign{\smallskip}\hline
\end{tabular}\\
\end{center}
\textbf{Note:} $\alpha_{\rm CE}$ = CE ejection parameter; $n(q)$ =
initial mass ratio distribution; ecc = eccentricity distribution
of binary orbit; $Z$ = metallicity.
\begin{center}
\end{center}
\end{table}

   \begin{figure*}
   \centering
   \includegraphics[width=150mm,height=150mm,angle=270.0]{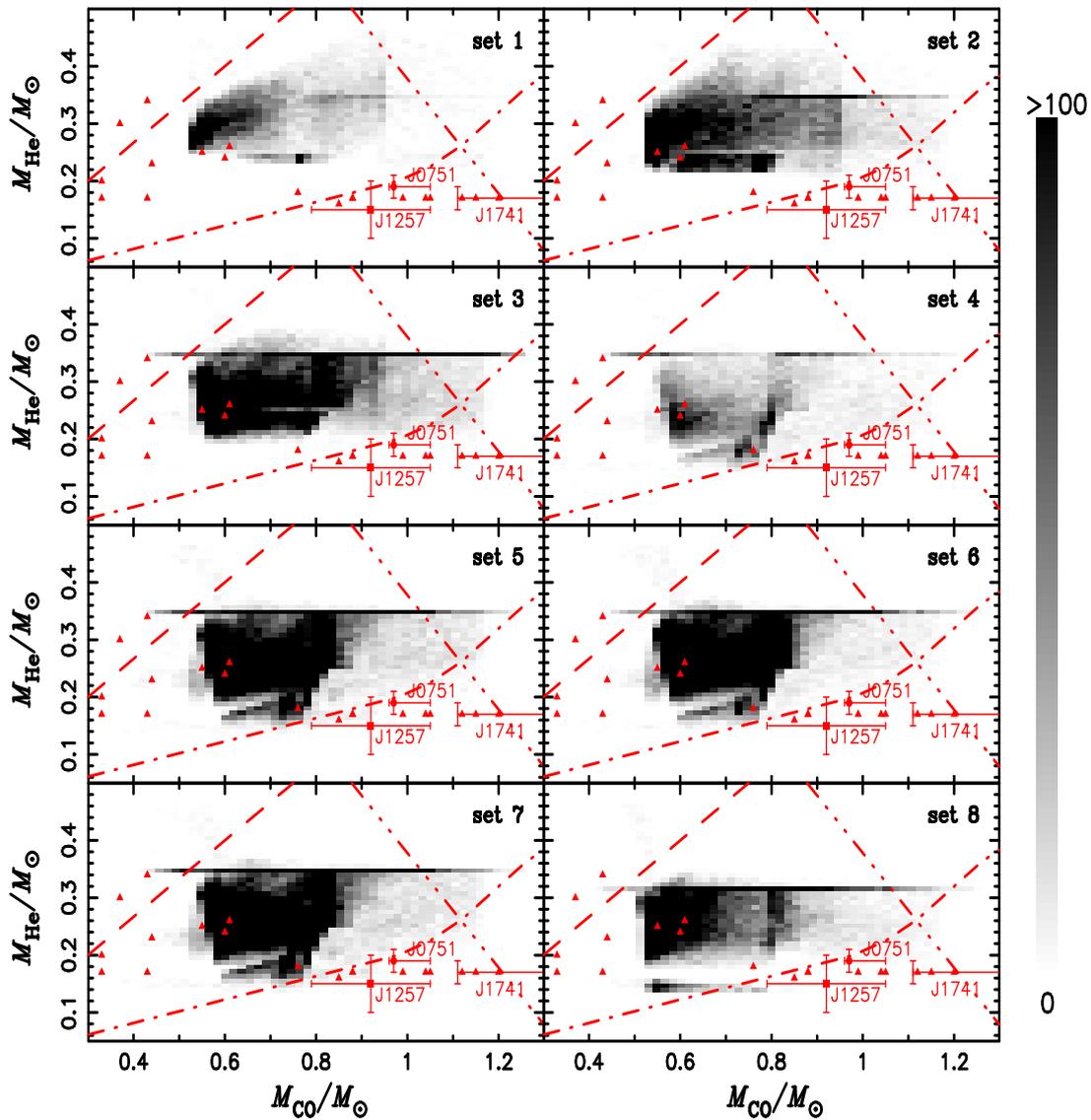}
   \caption{Distribution of CO WD and He WD masses for different BPS sets.
   The mass transfer between the CO WD and the He WD is always unstable for the systems above the dashed line ($q=M_{\rm He}/M_{\rm CO}>2/3$),
   while it is always stable below the dot-dashed line (Marsh et al. \cite{MARSH04}). The systems located at the right side of the
   triple-dot-dashed line have a mass of $\geq1.378$ $M_{\odot}$. Three extremely low-mass (ELM) WD systems
   (J0751, J1741 and J1257+5428), whose component masses were accurately measured,
   are clearly located in the region allowing stable mass transfer (Kulkarni \& van
   Kerkwijk \cite{KULKARNI10};
   Marsh et al. \cite{MARSH11};
   Kilic et al. \cite{KILIC14}). The filled triangles represent the systems from the ELM survey that may merge within 10 Gyr, where
   $i=60^{\rm \circ}$ is assumed for systems with unknown inclinations (Kilic et al. \cite{KILIC12}).}

              \label{Fig2}%
    \end{figure*}

\section{Results}\label{sect:3}
If a CO + He WD pair is the progenitor of a CRT, its merger should
belong to an old population, and the merging rate should also
match the birth rate of the CRTs derived from observations. The
delay time of the mergers is a function of the masses of the
components, as well as the orbital period, which is itself a
function of the masses of the components during CE evolution, as
is the mass ratio of the components. Therefore we identified
possible progenitors of the CRTs by the masses of its components.
An additional criterion was that, in theory, the CRTs have a
tendency to result from low-mass WD pairs (Shen et al.
\cite{SHENK10}; Waldman et al. \cite{WALDMAN11}), which was also
considered when identifying possible progenitors.

\subsection{Mass ratio?}\label{sect:3.1}
\begin{figure}
   \centering
   \includegraphics[width=80mm,height=80mm,angle=270.0]{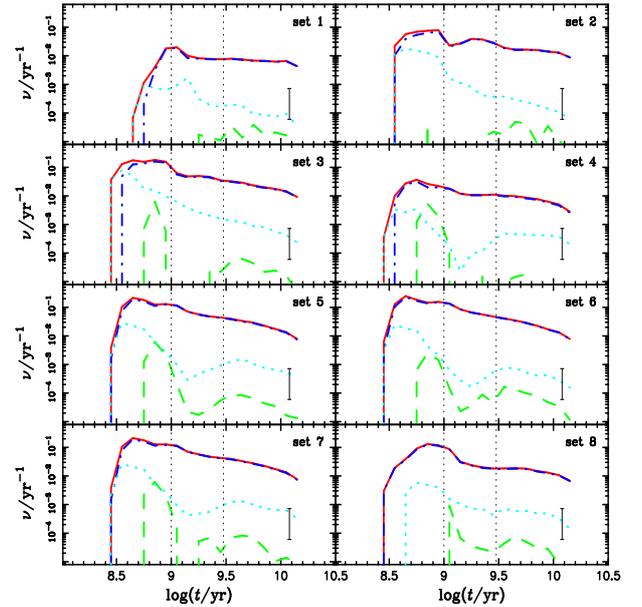}
   \caption{Evolution of the birth rate of potential candidates of
CRT progenitors. The red solid line is for all the potential
systems, while the green dashed and light-blue dotted lines are
for those with always unstable and stable mass transfer,
respectively. The blue dot-dashed line represents the systems
whose mass transfer is neither always stable nor always unstable
(see also Fig. \ref{Fig1}). The vertical bar indicates the range
of the birth rate of CRTs, normalized to the birth rate of SNe Ia
in S0/E galaxies (Mannucci et al. \cite{MANNUCCI05}). The two
vertical dotted lines mark the positions of delay times of 1 Gyr
and 3 Gyr.}
\label{Fig3}%
    \end{figure}

Figure \ref{Fig2} shows the distributions of $M_{\rm He}$ and
$M_{\rm CO}$ of the systems that are potential candidates for CRT
progenitors, and Fig. \ref{Fig3} presents the evolution of the
merging rate of the potential progenitors for different mass-ratio
constraints. At first glance, the distributions in Fig. \ref{Fig2}
are all similar, that is, most of the systems have a CO WD of
0.55-0.85 $M_{\odot}$ and a He WD of 0.2-0.35 $M_{\odot}$.
Notably, in every set, there is a concentration of He WDs at
$\sim0.35$ $M_{\odot}$ ($\sim0.32$ $M_{\odot}$ for the case of
$Z=0.02$), which result from quenched helium stars, meaning that
the mass-transfer between a CO WD and a low-mass helium star
begins when the helium star is still a helium main-sequence star
on a short orbital period, and then the helium burning is quenched
when the mass of the helium star reaches its minimum mass for
helium burning (see Han et al. \cite{HAN02}), which leads to the
formation of a He WD of $\sim0.35$ $M_{\odot}$ ($\sim0.32$
$M_{\odot}$ for the case of $Z=0.02$). See also Fig. \ref{Fig1}
for subchannels for Cases Ba, Ca, and Da.

From Fig. \ref{Fig2}, we can see that most of the systems are
located between the dashed and dot-dashed lines. The systems where
$q=M_{\rm He}/M_{\rm CO}>2/3$, whose mass transfer is always
dynamically unstable, are rare and typically have a delay time
shorter than 1 Gyr (see the dashed line in Fig. \ref{Fig3}). In
addition, the birth rate of mergers of such a high mass ratio is
also much lower than the birth rate of CRTs. Thus, it can be said
at the very least that not all WD pairs undergoing unstable mass
transfer contribute to the total number of CRTs, and that the
number of CRTs they produce is negligible, if they produce CRTs at
all. For systems below the dot-dashed line, the mass-transfer is
always stable. Such systems usually have a CO WD with a mass
higher than 0.7 $M_{\odot}$. Although the birth rate of mergers
from such systems matches those of CRTs  with a delay time longer
than 10 Gyr (the dotted line in Fig. \ref{Fig3}), there should be
many CRTs from young populations if these systems are indeed CRT
progenitors. Moreover, in theory, a detonation at the bottom of a
helium shell surrounding a relatively massive CO WD can result in
a second detonation, the expected result of which would be a SN Ia
(Shen \& Bildsten \cite{SHENK14}; Kilic et al. \cite{KILIC14}).
Furthermore, the material produced by the thermonuclear explosion
is mainly $^{\rm 56}$Ni, the spectra of which cannot match the
unique features of CRTs. From these perspectives, it seems
impossible for such systems to be the progenitors of CRTs.

It can be seen from both Figs. \ref{Fig2} and \ref{Fig3} that the
systems whose mass transfer processes are neither always stable
nor always unstable dominate the population of CO + He WD pairs.
Whether the mass transfer is dynamically stable or not depends on
the synchronization timescale $\tau_{\rm s}$ of the accretor, that
is, dynamically unstable mass transfer is more likely for a long
$\tau_{\rm s}$ (Marsh et al. \cite{MARSH04}). Here, $\tau_{\rm s}$
may be the manifestation of a torque resulting from dissipative
coupling, tidal or magnetic, which may be different between
different CO + He WD pairs. However, the merging rate of these
system is much higher than the birth rate of CRTs, and too many
mergers are young, which cannot account for the trend of CRTs
appearing among old populations (Fig. \ref{Fig3}). In addition,
most of the systems have a CO WD of larger than 0.6 $M_{\odot}$,
on which the helium detonation leaves mainly $^{\rm 56}$Ni and
unburnt helium, and the predicted spectrum from such explosion is
unlikely to fit the unique features of CRTs (Waldman et al.
\cite{WALDMAN11}).

In Fig. \ref{Fig2}, we also plot some WD pairs from the ELM
survey, which may merge within 10 Gyr (assuming $i=60^{\rm \circ}$
for the systems with unknown inclinations, Kilic et al.
\cite{KILIC12}). The distribution of the sample seems to favor
low-mass WDs compared with our calculation, which is very likely
due to an observational selection effect, since the effect of
cooling during evolution on the brightness of the WDs is not
considered here, and the aim of the project is to search extremely
low-mass WDs (Kilic et al. \cite{KILIC11}). However, in Fig.
\ref{Fig2}, we can see that many systems with an extremely
low-mass WD from the ELM survey are located in the region where
mass transfer is always stable, notably J0751, J1257+5428 and
J1741. These systems are suspected to be progenitors of
subluminous SNe Ia via double detonation (Branch et al.
\cite{BRA95}; Solheim \& Yungelson \cite{SOLHEIM05}; Hillebrandt
et al. \cite{HILLEBRANDT13}). If these systems produce SNe Ia,
they should contribute several percent of all SNe Ia based on the
calculation above, which is higher than the estimate of 1\% by
Solheim \& Yungelson (\cite{SOLHEIM05}). Three main reasons
contribute to this difference. Generally, the CO WD should be more
massive than 0.8 $M_{\odot}$ if the second detonation is expected
in the center of the CO WD (Shen et al. \cite{SHENK10}), but there
is no such constraint in this paper. Secondly, we did not consider
the effect of cooling on the brightness of the WDs, which means
that early-formed high-mass WDs may be too dim to be discovered
(Fontaine et al. \cite{FON01}). Finally, the CO WD may receive the
angular momentum of the accreted material and spin up. The effect
of rapid rotation reduces the violence of helium flashes on the
surface of the CO WD (Yoon \& Langer \cite{YOON04}), and a helium
nova is expected instead of a helium detonation. At present, there
are clearly too many uncertainties about the double-detonation
model leading to subluminous SNe Ia, and we do not discuss it
further.

In Fig. \ref{Fig2}, we also plot a triple-dot-dashed line, at
which the total mass of a CO + He WD pair is equal to 1.378
$M_{\odot}$ (close to the Chandrasekhar mass limit, Nomoto et al.
\cite{NTY84}). The systems located at the right side of the line
could produce normal SNe Ia if helium is stably burned into carbon
and oxygen to increase the mass of the CO WD. However, according
to the study of Shen et al. (\cite{SHEN12}), one may expect that
after the destruction of the less massive He WD, a giant-like
structure can be formed during the accretion stage. As a result of
the relatively long time during which the giant-like structure is
maintained, one could expect to lose about half a solar mass from
the system, and thus it should be very difficult for the CO WD to
increase its mass to 1.378 $M_{\odot}$. Regardless of the exact
mass, these systems are still interesting candidates as possible
progenitors of SNe Ia, and numerical simulations of the merging of
such systems should be studied. Irrespective of the uncertainties
during the merging stage, these systems cannot be the progenitors
of CRTs for the relatively massive CO WDs (see Fig. \ref{Fig2}).

Based on this discussion, it seems impossible to distinguish
progenitors of CRTs from other CO + He WD pairs by constraining
the mass ratio alone .

   \begin{figure}
   \centering
   \includegraphics[width=80mm,height=80mm,angle=270.0]{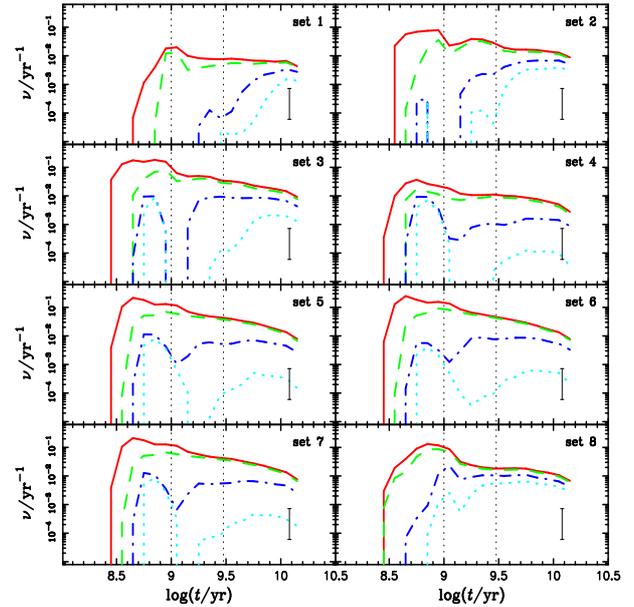}
   \caption{Evolution of the birth rate of the merger of CO WD + He WD pairs
   for different CO WD mass constraints. The red solid line is for all WD pairs
   and the green dashed, blue dot-dashed and light-blue dotted lines are for the pairs with a CO WD
   of $\leq0.8$ $M_{\odot}$, $\leq0.6$ $M_{\odot}$, and $\leq0.55$ $M_{\odot}$,
   respectively. Two vertical dotted lines mark the delay times
of 1 Gyr and 3 Gyr.} \label{Fig4}%
    \end{figure}

\subsection{CO WD mass?}\label{sect:3.2}
In Fig. \ref{Fig4}, we plot the evolution of the merging rate of
CO + He WD pairs for different CO WD mass constraints. From the
figure, we can see a significant trend that the shortest delay
time increases with decreasing CO WD mass constraint, which is
expected because the shortest delay time is determined by the
evolutionary timescale of the primordial secondary. A lower CO WD
mass means a less massive primordial primary and a less massive
primordial secondary, which in turn results in a long evolutionary
timescale of the secondary. From the figure, we can see that the
mergers with long delay times are dominated by the pairs with CO
WDs of larger than 0.6 $M_{\odot}$ due to a relatively longer
$t_{\rm GW}$. In addition, for $M_{\rm CO}\leq0.6$ $M_{\odot}$,
the mergers seem to be divided into two age groups, one that is
mainly younger than 1 Gyr, the other older than 1 Gyr. The reason
they form these two groups is that they are derived from different
evolutional channels: the young group mainly from the 2RLOF + 2CE
channel, the old group mainly from the 2CE + 1RLOF channel. For
the second group ($M_{\rm CO}\leq0.55$ $M_{\odot}$), both the
delay time and the merging rate seem to be consistent with the
constraints from CRT observations. However, for the set in which
Z=0.02, even under the constraint of $M_{\rm CO}\leq0.55$, the
merging rate at an age older than 10 Gyr is still much higher than
the birth rate of CRTs. This is mainly due to the effect of
metallicity on the initial-to-final mass relation (IFMR), which
stipulates that a higher metallicity leads to a lower WD mass for
a star with a given initial mass (Han et al. \cite{HAN94}; Umeda
\& Nomoto \cite{UN99}; Meng et al. \cite{MENG08}). So, relative to
the low-metallicity sample, the percentage of low-mass WDs from
the high-metallicity sample is higher. In any case, it also seems
impossible to distinguish the progenitor of CRTs from other CO +
He WD pairs by constraining the CO WD mass alone.

   \begin{figure}
   \centering
   \includegraphics[width=80mm,height=80mm,angle=270.0]{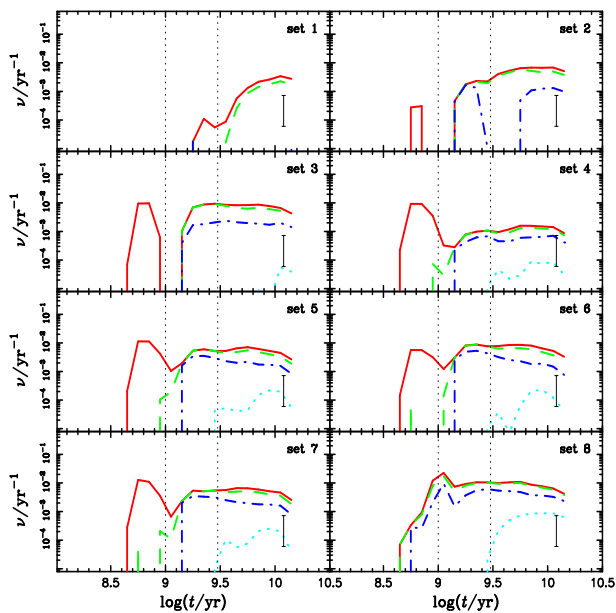}
   \caption{Evolution of the merging rate for systems with $M_{\rm CO}\leq0.6$
   $M_{\odot}$ (0.55 $M_{\odot}$ for set 8)
   and $M_{\rm He}\leq0.4$ $M_{\odot}$ (red solid line), $M_{\rm He}\leq0.3$ $M_{\odot}$ (green dashed line),
   $M_{\rm He}\leq0.25$ $M_{\odot}$ (blue dot-dashed line), and $M_{\rm He}\leq0.2$ $M_{\odot}$ (light-blue dotted line).
   The two vertical dotted lines indicate the delay times of 1 Gyr and 3 Gyr.}
   \label{Fig5}%
    \end{figure}
\subsection{Low-mass CO WD and low-mass He WD?}\label{sect:3.3}
Shen et al. (\cite{SHENK10}) and Waldman et al. (\cite{WALDMAN11})
indicated that if the CO WD mass is higher than 0.6 $M_{\odot}$,
helium detonation leaves mainly $^{\rm 56}$Ni and unburnt He, and
the predicted spectrum from this type of detonation is unlikely to
match the unique features of CRTs. We also found that a
significant proportion of the systems with a CO WD more massive
than 0.6 $M_{\odot}$ will merge within a delay time shorter than 3
Gyr, and the merging rate at a delay time of more than 10 Gyr is
much higher than that of CRTs. In addition, although the mass
ratio could be a very important factor in deciding the final
result of the merger, the mass ratios of CRT progenitors are
statistically indistinguishable from those of the general CO + He
WD binary population, as discussed in Sect. \ref{sect:3.1}. In
this section, we therefore only consider systems with a CO WD of
less mass than 0.6 $M_{\odot}$, and identify the progenitors of
CRTs by constraining the He WD mass. In Fig. \ref{Fig5}, we plot
the evolution of the merging rate for the systems with $M_{\rm
CO}\leq0.6$ and various He WD mass constraints. We again see the
trend that the shortest delay time increases with decreasing He WD
mass constraints, which is because the delay time is determined by
the evolutionary timescale of the primordial secondary: a low He
WD mass means a less massive primordial secondary, and,
consequently, a longer evolutionary timescale. The figure shows
that the merging rate at a delay time of longer than 10 Gyr mainly
arises from the systems with a He WD of $M_{\rm He}>0.25$
$M_{\odot}$. For the sets with $Z=0.001$, all the systems with a
He WD of $M_{\rm He}\leq0.25$ fulfill the loose delay-time
criterion, while almost all the systems with a He WD of $M_{\rm
He}\leq0.2$ fulfill the strict delay-time criterion. However,
although the systems with a He WD of $M_{\rm He}\leq0.2$ from sets
3, 4, 5, 6, and 7 match the birth rate of CRTs at a delay time
longer than 10 Gyr, there is not a single system with a He WD less
massive than 0.2 $M_{\odot}$ for set 2 and for set 1, even the
systems with the He WDs of $M_{\rm He}\leq0.25$ are rare (see also
Fig. \ref{Fig2}). The reason for this phenomenon is that for a low
$\alpha_{\rm CE}\lambda$, a system is more likely to merge during
the CE evolution, especially for systems with low-mass components
(see also Eq. \ref{eq:af}). However, the birth rate of mergers
with $M_{\rm He}\leq0.25$ at a delay time longer than 10 Gyr is
still higher than the observed birth rate of CRTs for set 2 and
$M_{\rm He}\leq0.3$ for set 1. Considering that the results from a
variable $\lambda$ are similar to that from a constant value of
$\lambda=1.0$ (Claeys et al. \cite{CLAEYS14}), we assume that
$\alpha_{\rm CE}\lambda\geq0.5$, that is, $\alpha_{\rm
CE}\geq0.5$, is reasonable since the thermal energy is not
included in our simulations. In conclusion, based on the results
in this figure and the numerical simulations from Shen et al.
(\cite{SHENK10}) and Waldman et al. (\cite{WALDMAN11}), it is
necessary that the CO WD is less massive than 0.6 $M_{\odot}$ and
the He WD is less massive than 0.25 $M_{\odot}$ for WD pairs to
fulfill the constraints from both the host population age and the
birth rate of CRTs. These systems could be the progenitors of
CRTs. This is also consistent with the numerical simulation by
Waldman et al. (\cite{WALDMAN11}). However, these constraints
could be $M_{\rm CO}\leq0.55$ and $M_{\rm He}\leq0.2$ $M_{\odot}$
for a population with a metallicity of $Z=0.02$ as a result of the
influence of metallicity on the IFMR (Meng et al. \cite{MENG08}).

\section{Discussion}\label{sect:4}
We identified the progenitors of CRTs from a general binary
population by constraining their component masses to fulfill the
age and birth rate constraints of CRTs. However, many questions
still remian open, such as the age of CRTs, the ejecta mass of the
explosion, and the explosion mechanism itself. We discuss these
problems below.

\subsection{Uncertainties}\label{sect:4.1}
As shown in Figs. \ref{Fig2}, \ref{Fig4}, and \ref{Fig5}, among
all the input parameters in the BPS simulations, $\alpha_{\rm
CE}\lambda$, that is, $\alpha_{\rm CE}$ and $\lambda$, is the most
important parameter to affect the BPS results. The low-mass
systems become rarer with the decrease of $\alpha_{\rm CE}\lambda$
because of the merging of the low-mass systems during the CE
evolution for a low value of $\alpha_{\rm CE}\lambda$ (see sets 1,
2, 3, and 5 in Figs \ref{Fig2}, \ref{Fig4}, and \ref{Fig5}).
However, at present, the values of $\alpha_{\rm CE}$ and $\lambda$
are quite uncertain (Han \cite{HAN98}; Politano \& Weiler
\cite{POLITANO07}; Xu \& Li \cite{XL10}; Zorotovic et al.
\cite{ZOROTOVIC10}; Zuo \& Li \cite{ZUO14}; Claeys et al.
\cite{CLAEYS14}). Our results are based on the assumption that
$\alpha_{\rm CE}\lambda\geq0.5$. At present, we cannot exclude the
possibility of $\alpha_{\rm CE}\lambda<0.5$, although such a low
value seems to be unreasonable since the thermal energy is not
included for CE evolution in our BPS simulation. If $\alpha_{\rm
CE}\lambda$ were lower than 0.5, e.g. 0.25, the progenitors of
CRTs could be the systems with $M_{\rm CO}\leq0.6$ and $M_{\rm
He}\leq0.3$ $M_{\odot}$ (set 1 in Fig. \ref{Fig5}). Our other
conclusions are not significantly affected by the value of
$\alpha_{\rm CE}\lambda$ (see the following discussions).

\subsection{Age}\label{sect:4.2}
Although the observations made to constrain the progenitor
population of the CRTs favored an old population (Kasliwa et al.
\cite{KASLIWA12}; Yuan et al. \cite{YUANF13}; Lyman et al.
\cite{LYMAN13}), the possibility is still not completely excluded
that a small part of CRTs might be hosted by a relatively young
population, such as SN 2001co (Lyman et al. \cite{LYMAN13}). In
addition, quantitatively, the association of CRTs with the regions
of ongoing star formation in their host galaxies matches that of
SNe Ia (Lyman et al. \cite{LYMAN13}), which implies that some CRTs
may originate from relatively young populations. There might be an
observational selection effect because a subluminous supernova is
more easily discovered in the border region than in the inner
region of a host galaxy. From Fig. \ref{Fig5}, we know that some
pairs with low-mass CO WDs and low-mass He WDs have a medium delay
time (between 1 Gyr and 3 Gyr), although there are relatively few
of them. At present, the sample of CRTs is still small, and the
statistical uncertainties about the population of the progenitors
of CRTs are very large. This means that additional observations
are necessary to confirm whether all the CRTs are hosted by old
populations.

   \begin{figure}
   \centering
   \includegraphics[width=80mm,height=80mm,angle=270.0]{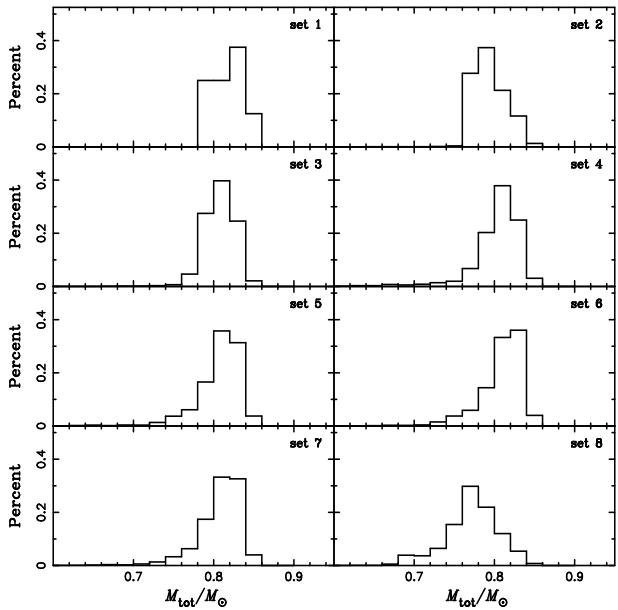}
   \caption{Distribution for different sets of the total mass of the WD pairs that qualify under the constraints
   that both are hosted by an old population and match the birth rate of CRTs.}
   \label{Fig6}%
    \end{figure}

\subsection{Ejecta mass}\label{sect:4.3}
The ejecta mass of CRTs is concentrated at 0.4-0.7 $M_{\odot}$
(Kasliwa et al. \cite{KASLIWA12}), while that of a small minority
of CRTs is beyond this range (Perets et al. \cite{PERETS10};
Valenti et al. \cite{VALENTI14}). The scatter of the peak
luminosity of the CRTs is small, 0.5 mag. These results could
indicate that the distribution of the parameter determining the
peak luminosity have a narrow peak with a wing, and the parameter
could be relevant with one of the masses related to the low-mass
CO + He WD pairs. Clearly, for the systems satisfying both the old
population and the birth rate constraints, neither the CO WD mass
nor the He WD mass fulfills the constraints of the ejecta mass. An
ideal candidate for the parameter is the total mass of the merger.
In Fig. \ref{Fig6}, we present a histogram of the total mass of
the WD pairs qualifying for the above constraints. The figure
shows that the distribution of the total mass of the mergers has a
narrow peak with a wing, although the position of the peak is
dependent of the initial conditions of BPS set. However, the total
mass is significantly higher than the ejecta mass of CRTs. This
implies that if the low-mass CO + He WD pairs are the progenitors
of CRTs, the merger probably is not completely destroyed, and a
remnant survive after the explosion.

In theory, it is generally believed that a He-shell detonation is
followed by a second detonation in the core of the CO WD, although
it is still not clear whether and where the second detonation will
occur (see Fink et al. \cite{FINK07}; Kromer et al.
\cite{KROMER10}). Shen et al. (\cite{SHENK10}) investigated
He-shell detonations for three different CO WD masses (0.6, 1.0
and 1.2 $M_{\odot}$) with different He-shell masses (0.02, 0.05,
0.1, 0.2, and 0.3 $M_{\odot}$), and Waldman et al.
(\cite{WALDMAN11}) extended the work of Shen et al.
(\cite{SHENK10}) to low-mass CO WDs (0.45-0.6 $M_{\odot}$) with a
0.2 $M_{\odot}$ He shell. Similar to Waldman et al.
(\cite{WALDMAN11}), Sim et al. (\cite{SIM12}) also studied the
He-shell detonation on a low-mass CO WD, but extended the
simulation to the cases where a second detonation occurs. Based on
these studies, if the second detonation occurs, the models show a
brighter light curve than those from observations (Sim et al.
\cite{SIM12}). If the second detonation does not occur, the peak
luminosity is similar to, but the evolution of the light curve is
much faster and the ejecta mass is less massive than, those from
observations (Waldman et al. \cite{WALDMAN11}). Notably, even the
best-fit model to the data of SN 2005E in Waldman et al.
(\cite{WALDMAN11}) is still somewhat fainter at peak, and clearly
fades more rapidly after maximum light than, the observed light
curve of SN 2005E. The observed ejecta mass is also too high for a
shell detonation, which suggests that the CO core also
participated in the explosion. At the same time, as indicated by
the distribution of the total mass of CO + He WD pairs in Fig.
\ref{Fig6}, to account for the ejecta mass of CRTs, a remnant
should survive the thermonuclear explosion if low-mass CO + He WD
pairs are progenitors of CRTs. This seems to require a certain
level of fine-tuning if the properties of CRTs are to be matched
by exploding mergers. Even though many theoretical works have
addressed the CRT model, more efforts may still be necessary to
obtain a conclusive result on this, especially regarding whether
the second detonation is triggered, and how a remnant is left.
Since the second detonation in the center of the CO WD usually
destroys the whole merger, an off-center detonation might be an
alternative (Shen \& Bildsten \cite{SHENK14}).

\subsection{Other mechanisms}\label{sect:4.4}
Kasliwa et al. (\cite{KASLIWA12}) argued that CRTs cannot be the
standard result of a thermonuclear explosion of a CO WD, which SNe
Ia are, and presented two main differences from SNe Ia to support
their argument. One is that no CRT conforms to the Philips
relation (Philips \cite{PHI93}), and the other is that Fe-group
elements seen in all SNe Ia are absent from the nebular spectra of
CRTs. Kasliwa et al. (\cite{KASLIWA12}) examined the predictions
from several theoretical models such as the deflagration of a
sub-Chandrashekar-mass WD (Woosley \& Weaver \cite{WOOSLEY94}) and
the accretion-induced collapse of a rapidly rotating ONeMg WD into
a neutron star (Metzger et al. \cite{METZGER09}). They found that
none of their predictions is consistent with the properties of
CRTs. Yuan et al. (\cite{YUANF13}) suggested that one possible
interpretation of the unusual subluminous CRTs is that it might be
the result of a violent merger of two equal-mass CO WDs (Pakmor et
al. \cite{PAKMOR10}). However, such a violent merger has either a
very short delay time or a very low birth rate (see Fig. 1 in Meng
et al. \cite{MENGXC11}), and the ejecta mass resulting from such a
violent merger is too massive compared with the observed ejecta
masses. In addition, it could be difficult for the merger of two
equal-mass WDs to reproduce the helium line in the spectra of some
CRTs (Kasliwa et al. \cite{KASLIWA12}). Another possible mechanism
is a variation of the violent merger model, that is, a violent
merger from a CO + He WD pair, which was suggested to explain the
existence of subluminous SNe Ia (Pakmor et al. \cite{PAKMOR13}).
However, this model also fails to explain the observed CRT ejecta
masses. Moreover, observationally, it could be difficult to
distinguish the difference between the classical double-detonation
model and the violent merger model from the CO + He WD pairs,
especially for the delay time and the birth rate.

One of the CRTs (PTF 09dav) in the sample of Kasliwa et al.
(\cite{KASLIWA12}) exhibited a hydrogen line in its nebular
spectra, and Kasliwa et al. (\cite{KASLIWA12}) suggested that
there might have been a series of nova eruptions prior to this
particular CRT to account for this feature (see also Shen et al.
\cite{SHENK13}). In fact, there is another possible explanation
for the hydrogen line, which is that the CRT with the hydrogen
line in its nebular spectra evolved from the Case Db subchannel,
which is similar to the proposition of Livio \& Riess
(\cite{LR03}) when explaining the properties of SN 2002ic. If this
CRT results from this subchannel, one would expect a fairly thin
shell of hydrogen ejecta, similar to what is seen in planetary
nebulae with close binary cores. Judging from the estimation in
Kasliwa et al. (\cite{KASLIWA12}), and given the velocity of 10
${\rm km}$ ${\rm s}^{\rm -1}$ for the ejecta shell, the CE
ejection should cease 7800 yr before the supernova if the shell is
photoionized, and 260 yr before the supernova if the shell is
collision-ionized when the shock front reaches the shell.

\section{Conclusions}\label{sect:5}

In summary, we tried to investigate whether and which types of
low-mass CO + He WD pairs can result in CRTs by examining whether
the results of their evolution matches the host population ages
and birth rates of CRTs in general. We presented four channels
that produce CO + He WD pairs and showed that the classical
channel to produce WD + WD pairs, in which a WD + WD pair only
experiences two CE phases and does not undergo a stable RLOF, is
not the dominant channel. This means that future attention should
be paid to the stable RLOF when studying the formation of WD
pairs. We found that the CO + He WD pairs with CO WDs less massive
than 0.6 $M_{\odot}$ and He WDs less massive than 0.25 $M_{\odot}$
may qualify as CRT progenitors under the constraints of both host
population ages and birth rates if $\alpha_{\rm
CE}\lambda\geq0.5$. If $\alpha_{\rm CE}\lambda$ were lower than
0.5, for example 0.25, the upper limit for the mass of the He WDs
could be relaxed to 0.3 $M_{\odot}$. However, the helium WD mass
is lower than, while the total mass of the CO + He WD pairs is
higher than, the ejecta mass of CRTs. If these low-mass WD pairs
are the progenitors of CRTs, the CO WD should participate in the
explosion, and a binding remnant could remain after the explosion.
Clearly, many problems regarding CRTs need to be addressed both in
theory and observation, especially to determine whether the second
detonation is ignited at the center of a CO WD following a helium
detonation at the surface of the CO WD, and whether there is a
bound remnant after the CRT explosion, and whether all CRTs are
from an old population. Finally, we suggest that the CRT with a
hydrogen line in its nebular spectra might have evolved from the
Case Db channel.

\begin{acknowledgements}
We are grateful to the anonymous referee for his/her comments that
improved this manuscript. This work was partly supported by NSFC
(11473063,11390374) and Key Laboratory for the Structure and
Evolution of Celestial Objects, Chinese Academy of Sciences. Z.H.
thanks the support by the Strategic Priority Research Program
``The Emergence of Cosmological Structures" of the Chinese Academy
of Sciences, Grant No. XDB09010202.
\end{acknowledgements}

\end{document}